\documentclass{article}
\usepackage{spconf}
\usepackage{microtype}
\usepackage{float}
\usepackage{amsmath,amssymb,amsfonts}
\usepackage{pifont}
\usepackage[utf8]{inputenc}
\usepackage[T1]{fontenc}
\usepackage{url}
\usepackage[pdfusetitle]{hyperref}
\usepackage[capitalize,nameinlink]{cleveref}
\usepackage[nopostdot,nogroupskip,nonumberlist,toc,numberline,acronyms]{glossaries}
\usepackage{algorithmic}
\usepackage{graphicx}
\usepackage[dvipsnames]{xcolor}
\usepackage[inline]{enumitem}
\usepackage{booktabs}
\usepackage{makecell}
\usepackage{mathtools}
\usepackage{xparse}
\usepackage{siunitx}
\usepackage{tikz}
\usepackage{fdsymbol}
\usepackage{relsize}
\usepackage[super]{nth}
\usepackage[style=ieee,backend=biber]{biblatex}

\usetikzlibrary{external,shapes,arrows,positioning,intersections,fit,calc}
\usetikzlibrary{trees,mindmap,patterns.meta,arrows.meta,colorbrewer,shapes.geometric}
\usetikzlibrary{perspective,3d,backgrounds,decorations.pathreplacing}

\hypersetup{hidelinks}

\renewbibmacro*{doi+eprint+url}{%
  \iftoggle{bbx:doi}
    {\iffieldundef{doi}
       {\iftoggle{bbx:eprint}
          {\iffieldundef{eprint}
             {\iftoggle{bbx:url}
                {\usebibmacro{url+urldate}}
                {}}
             {\usebibmacro{eprint}}}
          {\usebibmacro{eprint}}}
       {\printfield{doi}}}
    {}%
}

\AtEveryBibitem{
    \clearfield{urlyear}
    \clearfield{urlmonth}
    \clearfield{urlday}
    \clearfield{day}
    \clearfield{series}
    \clearfield{number}
    \clearfield{volume}
    \clearfield{volumes}
    \clearfield{pages}
    \clearfield{issn}
    \clearfield{isbn}
    \clearfield{note}
    \clearlist{publisher}
    \clearlist{language}
    \clearlist{location}
    \clearname{editor}
    \clearfield{doi}
    \clearfield{url}
}

\colorlet{lightblue}{Paired-A}
\colorlet{blue}{Paired-B}
\colorlet{lightgreen}{Paired-C}
\colorlet{green}{Paired-D}
\colorlet{lightred}{Paired-E}
\colorlet{red}{Paired-F}
\colorlet{lightorange}{Paired-G}
\colorlet{orange}{Paired-H}
\colorlet{lightpurple}{Paired-I}
\colorlet{purple}{Paired-J}
\colorlet{lightgray}{Greys-E}
\colorlet{gray}{Greys-J}

\newlist{inlineenum_abc}{enumerate*}{1}
\setlist[inlineenum_abc,1]{label=\emph{\alph*})}

\newlist{inlineenum_123}{enumerate*}{1}
\setlist[inlineenum_123,1]{label=\emph{\arabic*})}

\newlist{noindent_descr}{description}{1}
\setlist[noindent_descr,1]{wide}

\DeclareMathAlphabet{\mathmybb}{U}{bbold}{m}{n}

\def\Model{{\mathcal{M}}}
\def\Real{{\mathbb{R}}}
\def\Complex{{\mathbb{C}}}
\def\Bin{{\{0,1\}}}

\newcommand\norm[1]{\lVert#1\rVert}
\def\ttt{{\smash{(t)}}}
\def\uuu{{\smash{(u)}}}

\newcommand{\colordot}[1]{\textcolor{#1}{\raisebox{-1pt}{\scalebox{1.3}{$\bullet$}}}}

\newcommand{\ShowLen}[2]{\texttt{\string#1} & {\small\texttt{\the#1}} & #2\\}

\makeatletter
\renewcommand\section{\@startsection{section}{1}{\z@}%
    {-3.5ex \@plus -1ex \@minus -.4ex}%
    {2.3ex \@plus .2ex \@minus .1ex}%
    {\normalfont\Large\bfseries}}
\renewcommand\subsection{\@startsection{subsection}{2}{\z@}%
  {-3.25ex \@plus -1ex \@minus -.4ex}%
  {1.5ex \@plus .2ex \@minus .1ex}%
  {\normalfont\large\bfseries}}
\makeatother

\newacronym{dl}{DL}{deep learning}
\newacronym{dnn}{DNN}{Deep Neural Network}
\newacronym{cnn}{CNN}{Convolutional Neural Network}
\newacronym{rnn}{RNN}{Recurrent Neural Network}
\newacronym{lstm}{LSTM}{Long Short-Term Memory}
\newacronym{gru}{GRU}{Gated Recurrent Unit}
\newacronym{glu}{GLU}{Gated Linear Unit}
\newacronym{cv}{CV}{Computer Vision}
\newacronym{nlp}{NLP}{Natural Language Processing}
\newacronym{mac}{MAC}{Multiply-Accumulate Operation}
\newacronym{moe}{MoE}{Mixture of Experts}
\newacronym{uf}{UF}{Utilization Factor}
\newacronym{dynnn}{DynNN}{Dynamic Neural Network}
\newacronym{scalnn}{ScalNN}{Scalable Neural Network}
\newacronym{dyncp}{DynCP}{Dynamic Channel Pruning}
\newacronym{se}{SE}{Speech Enhancement}
\newacronym{sd}{SD}{Speech De-reverberation}
\newacronym{ss}{SS}{Source Separation}
\newacronym{aec}{AEC}{Acoustic Echo Cancellation}
\newacronym{sv}{SV}{Speaker Verification}
\newacronym{asr}{ASR}{Automatic Speech Recognition}
\newacronym{sed}{SED}{Sound Event Detection}
\newacronym{asc}{ASC}{Acoustic Scene Classification}
\newacronym{kws}{KWS}{Keyword Spotting}
\newacronym{dns}{DNS}{Deep Noise Suppression}
\newacronym{sqp}{SQP}{Speech Quality Prediction}
\newacronym{vad}{VAD}{Voice Activity Detection}

\newacronym{stft}{STFT}{Short-Time Fourier Transform}
\newacronym{pesq}{PESQ}{Perceptual Evaluation of Speech Quality}
\newacronym{stoi}{STOI}{Short-Time Objective Intelligibility}
\newacronym{snr}{SNR}{Signal-to-Noise Ratio}
\newacronym{sisdr}{SI-SDR}{Scale-Invariant Signal-to-Distortion Ratio}
\newacronym{sq}{SQ}{Speech Quality}
\newacronym{wer}{WER}{Word Error Rate}
\newacronym{eer}{EER}{Equal Error Rate}
\newacronym{mindcf}{minDCF}{Minimum Detection Cost Function}
\newacronym{rms}{RMS}{Root Mean Square}
\newacronym{r2}{$\smash{\cramped{\text{R}^2}}$}{Coefficient of Determination}
\newacronym{mae}{MAE}{Mean Absolute Error}
\newacronym{rmse}{RMSE}{Root Mean Square Error}
\newacronym{rocauc}{ROC-AUC}{Area Under the Receiver Operating Characteristic Curve}
\newacronym{mi}{MI}{Mutual Information}
\newacronym{vbd}{VB+D}{VoiceBank+DEMAND}
\newacronym{vb}{VB}{VoiceBank}
\newacronym{wccn}{WCCN}{Within-Class Covariance Normalization}
\newacronym{lda}{LDA}{Linear Discriminant Analysis}

\title{From Diet to Free Lunch: Estimating Auxiliary Signal Properties\\ using Dynamic Pruning Masks in Speech Enhancement Networks}

\name{
Riccardo Miccini$^{\flat \sharp}$,
Clément Laroche$^{\flat}$,
Tobias Piechowiak$^{\flat}$,
Xenofon Fafoutis$^{\sharp}$,
Luca Pezzarossa$^{\sharp}$
\thanks{This work has received funding from the European Union's Horizon research and innovation program under grant agreement No.\,101070374.}
}

\address{
$^\flat$\textit{GN Hearing} \quad
$^\sharp$\textit{Technical University of Denmark (DTU)}
}

\begin{document}
\topmargin=0mm
\ninept
\maketitle

\makeatletter
\def\@makefnmark{$\cramped{^{\smash{\@thefnmark}}}$}%
\makeatother

\glsunset{snr}
\glsunset{r2}

\begin{abstract}
\gls{se} in audio devices is often supported by auxiliary modules for \gls{vad}, \gls{snr} estimation, or \acrlong{asc} to ensure robust context-aware behavior and seamless user experience. 
Just like \gls{se}, these tasks often employ deep learning; however, deploying additional models on-device is computationally impractical, whereas cloud-based inference would introduce additional latency and compromise privacy.
Prior work on \gls{se} employed \gls{dyncp} to reduce computation by adaptively disabling specific channels based on the current input. 
In this work, we investigate whether useful signal properties can be estimated from these internal pruning masks, thus removing the need for separate models.
We show that simple, interpretable predictors achieve up to \qty{93}{\percent} accuracy on \gls{vad}, \qty{84}{\percent} on noise classification, and an \gls{r2} of \num{0.86} on $F_0$ estimation. 
With binary masks, predictions reduce to weighted sums, inducing negligible overhead. 
Our contribution is twofold: on one hand, we examine the emergent behavior of \gls{dyncp} models through the lens of downstream prediction tasks, to reveal what they are learning; on the other, we repurpose and re-propose \gls{dyncp} as a holistic solution for efficient \gls{se} and simultaneous estimation of signal properties.
\end{abstract}


\glsresetall[acronym]

\section{Introduction}
\label{sec:intro}

\gls{se} is a fundamental part of many devices aimed at improving communication, collaboration, or quality of life, such as hearing aids, audio wearables, and voice-activated systems.
Recent advances in \gls{dl} have achieved state-of-the-art performance in \gls{se}.
However, to ensure responsiveness, privacy, and offline operation, \gls{dl} models must be deployed directly on embedded devices, rather than relying on cloud-based inference.
The challenge of running \gls{dl} models for \gls{se} under tight computational constraints has spurred significant research on lightweight architectures such as 
PercepNet~\cite{valin_perceptually-motivated_2020,ge_percepnet_2022}, FullSubNet~\cite{hao_fullsubnet_2021}, and GTCRN~\cite{rong_gtcrn_2024}. 

While \gls{se} is often the main component in a speech processing pipeline, real-world solutions may include additional modules for \gls{sqp}, \gls{asc}, \gls{snr} estimation, or \gls{vad}. 
These auxiliary elements may be integrated in the \gls{se} model or inform downstream business logic, allowing a more seamless, robust, and tailored user experience.
For instance, instantaneous \gls{snr} or speech quality estimates may provide useful insights to end users, e.g., suggesting them to speak louder or to adjust the microphone arm when necessary.
Indeed, applications of this information to guide hybrid \gls{se} systems have been explored extensively: in \cite{lv_dccrn_2021,ge_percepnet_2022,schroter_deepfilternet_2023}, \gls{snr} estimates selectively enable parts of the model; in~\cite{zhang_vsanet_2023}, the authors jointly train for \gls{se} and \gls{vad}, achieving faster convergence and better performance; in~\cite{hsieh_tgif_2025}, speaker embeddings aid target speaker extraction within a small group of related users; similarly, other signal characteristics such as speaker gender, \gls{snr} level, or speech quality have proven useful in zero-shot model selection~\cite{sivaraman_sparse_2020,zezario_speech_2021}. 
While resource-efficient predictors have been investigated, such as those for \gls{sqp}~\cite{cumlin_dnsmos_2024, nilsson_resource-efficient_2024,nilsson_efficient_2025}, the computational overhead may still be substantial, even with non-\gls{dl} methods.

\begin{figure}[t]
    \centering
    \includegraphics[width=1.0\linewidth]{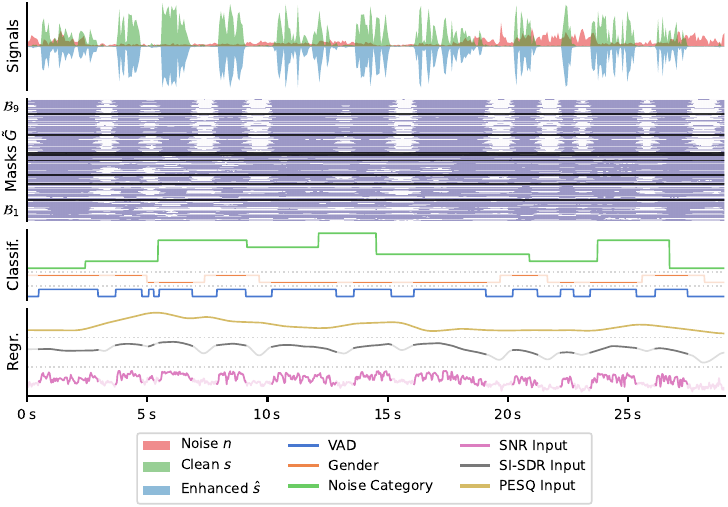}%
    \caption{Example generated data, showing clean/noise/enhanced audio, binary pruning masks, and a selection of ground truths; for relevant targets, we highlight regions with voice activity.}
    \label{fig:data_example}
\end{figure}

An orthogonal line of research is exploring \glspl{dynnn}, a class of models which adjust their computation depending on the current input~\cite{han_dynamic_2022}. 
Several techniques have been recently proposed in the context of speech processing, such as \cite{miccini_adaptive_2025,elminshawi_dynamic_2025,olsen_knowing_2025}.
Among these, \gls{dyncp}~\cite{jelcicova_peakrnn_2021,cheng_dynamic_2024,miccini_scalable_2025} involves skipping computation of some of the channels based on an input-dependent criterion, thus adaptively balancing efficiency and performance.
In \cite{miccini_scalable_2025}, this criterion is expressed by binary masks estimated using gating subnets inserted into each processing block of the \gls{se} backbone; see \cref{fig:data_example} for an example.
In their work, the authors empirically show that those masks seem to correlate with signal characteristics such as noise level or voice activity, raising the following questions: 
\begin{inlineenum_abc}
    \item what information is learned by the gating subnets?
    \item can we exploit it to perform those auxiliary tasks without needing dedicated models?
\end{inlineenum_abc}

In this article, we address these questions by drawing from recent literature investigating the latent representations of audio models~\cite{cumlin_impairments_2025,deng_investigating_2025}. 
In particular, we demonstrate how \gls{dyncp} gating subnets implicitly learn and encode relevant information about speech and acoustic conditions, despite being trained to efficiently allocate computational resources.
By using the binary pruning masks as input features for prediction models, we can perform tasks such as \gls{snr} estimation, \gls{sqp}, \gls{vad}, and estimation of other signal properties in real-time, at virtually no extra cost. 
We deliberately use linear/logistic regression to test whether the information is linearly accessible.
These models boast a straightforward formulation, inherently interpretable coefficients, and extremely low computational costs.
The latter is further enhanced by the use of binary inputs, which reduces dot-products to simpler gather-and-sum operations, similarly to~\cite{nilsson_resource-efficient_2024}.

Our perspective differs from previous work in three major ways: 
\begin{inlineenum_abc}
    \item instead of training dedicated, compute-heavy \gls{sqp} models on multiple targets~\cite{kumar_torchaudio-squim_2023,zhang_end--end_2021,zezario_deep_2023}, we exploit the \gls{dyncp} gating subnets as feature extractors;
    \item unlike \gls{se} systems where auxiliary cues are explicitly estimated and employed~\cite{sivaraman_sparse_2020,lv_dccrn_2021,ge_percepnet_2022,schroter_deepfilternet_2023,tan_speech_2021,zhang_vsanet_2023,hsieh_tgif_2025}, we show how we can extrapolate this information as a byproduct of \gls{dyncp}, suggesting similarities between \glspl{dynnn} and multi-task learning;
    \item prior work on audio embeddings focused on continuous representations~\cite{cumlin_impairments_2025,deng_investigating_2025}, while work on sparse activations did not tackle audio~\cite{srivastava_understanding_2015}.
\end{inlineenum_abc}
We therefore contribute the following:
\begin{inlineenum_123}
    \item a pipeline for computing and aligning \gls{dyncp} masks with auxiliary speech and acoustic attributes;
    \item performance evaluation across multiple classification, regression, and \gls{sv} tasks;
    \item extensive analyses and visualizations of the emergent behavior of \gls{dyncp} \gls{se} models, with the hope of encouraging further research in this area.
\end{inlineenum_123}

\section{Methods}
\label{sec:methods}

\begin{table}[t]
    \centering
    \caption{Overview of ancillary prediction tasks.}
    \label{tab:targets}
    \resizebox{1.0\linewidth}{!}{
    \begin{tabular}{@{} lccl @{}}
        \toprule
        \textbf{Name} & \textbf{Task type} & \textbf{Classes/IQR} & \textbf{Computed from} \\
        \midrule
        VAD & Classification & \num{2} & Clean energy envelope \\
        Gender & Classification & \num{2} & Metadata from \acrlong{vb}~\cite{veaux_voice_2013,yamagishi_cstr_2019} \\
        Accent & Classification & \num{6} & Metadata from \acrlong{vb}~\cite{veaux_voice_2013,yamagishi_cstr_2019} \\
        Noise category & Classification & \num{6} & Metadata from \acrshort{vbd}~\cite{valentini-botinhao_investigating_2016} and DEMAND~\cite{thiemann_demand:_2013} \\
        \midrule
        Input SNR & Regression & \qtyrange[range-units=single,range-phrase=--]{-13}{8}{\decibel} & \texttt{librosa.feature.rms} \\
        Enhanced SNR & Regression & \qtyrange[range-units=single,range-phrase=--]{1}{14}{\decibel} & \texttt{librosa.feature.rms} \\
        Input SI-SDR & Regression & \qtyrange[range-units=single,range-phrase=--]{-1}{10}{\decibel} & \texttt{auraloss.time.SISDRLoss} \\
        Enhanced SI-SDR & Regression & \qtyrange[range-units=single,range-phrase=--]{8}{17}{\decibel} & \texttt{auraloss.time.SISDRLoss} \\
        Input PESQ & Regression & \numrange[range-units=single,range-phrase=--]{1.2}{1.7} & \texttt{torch\_pesq.PesqLoss.mos} \\
        Enhanced PESQ & Regression & \numrange[range-units=single,range-phrase=--]{2.6}{2.9} & \texttt{torch\_pesq.PesqLoss.mos} \\
        $F_0$ & Regression & \qtyrange[range-units=single,range-phrase=--]{110}{200}{\hertz} & \texttt{pyworld.dio} and \texttt{pyworld.stonemask} \\
        \bottomrule
    \end{tabular}
    }
\end{table}

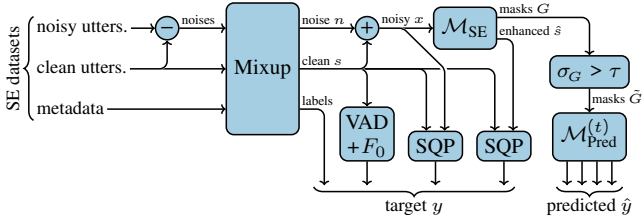
\begin{figure}[t]
    \centering
    \resizebox{1.0\linewidth}{!}{\begin{tikzpicture}[
    node distance=1em,
    font=\footnotesize,
    auto,
    scale=0.7,
]

\tikzset{every node/.append style={transform shape, align=center, text centered}}
\tikzset{strut/.style={execute at begin node=\vphantom{Q}}}
\tikzset{inputs/.style={strut, draw=none, inner sep=1pt, fill=none, font=\scriptsize}}
\tikzset{box/.style={rectangle, rounded corners=2pt, draw, inner sep=2pt, inner xsep=3pt, fill=lightblue}}
\tikzset{exbox/.style={box, text height=1.5ex, text depth=0.25ex, inner xsep=0pt}}
\tikzset{dot/.style={draw=none, fill=none, minimum width=0pt, minimum height=0pt, inner sep=0pt}}

\tikzset{ar/.style={draw, rounded corners=2pt, -{>[scale=0.5]}}}
\tikzset{arw/.style={draw=white, semithick, rounded corners=2pt, -, shorten >=0.1pt, shorten <=0.1pt}}
\tikzset{arnode/.style={strut, rounded corners=0pt, above right, at start, inner xsep=0.75pt, inner ysep=0.5pt, fill=none, font=\tiny}}

\tikzset{brace/.style={decorate,decoration={brace, mirror, raise=0pt, aspect=0.5}}}
\tikzset{bracenode/.style={strut, inner sep=0pt, fill=none, font=\scriptsize}}

\node [inputs] (x1) {noisy utters.};
\node [inputs, below right=0.75em and 0em of x1.south west] (s1) {clean utters.};
\node [inputs, below right=0.75em and 0em of s1.south west] (m1) {metadata};
\draw[brace] (x1.north west) -- (m1.south west) node[bracenode, rotate=90, midway, above=0pt of m1, yshift=4pt] {\gls{se} datasets};

\node [circle, draw, fill=lightblue, inner sep=0.5pt, right=1em of x1] (sub) {$-$};
\node [box, right=4em of s1.east, minimum height=5.5em] (mu) {Mixup};
\path (mu.north east)--(mu.south east) {
    coordinate (mu_out1) at (x1.east -| mu.east)
    coordinate (mu_out2) at (s1.east -| mu.east)
    coordinate (mu_out3) at (m1.east -| mu.east)
};

\matrix [below right=2em and 1em of mu.south east, column sep=0.5em, inner sep=0em, nodes=draw] {
    \node[dot] (t1) {}; & 
    \node[dot, minimum width=1.5em] (t2) {}; & 
    \node[dot, minimum width=1.5em] (t3) {}; &
    \node[dot, minimum width=1.5em] (t4) {}; &
    \node[dot, minimum width=2.5em] (t5) {}; \\
};
\draw[brace] ($(t1.south)!-4pt!(t4.south)$) -- ($(t4.south)!-4pt!(t1.south)$) node[bracenode, midway, below=4pt] (ylbl) {target $y$};

\node [exbox, above=of t2, minimum width=2.25em, text height=3.75ex] (vad) {VAD\\+ \!$F_0$};
\node [exbox, above=of t3, minimum width=2.25em] (sqp1) {SQP};
\path (sqp1.north west)--(sqp1.north east) {
    coordinate[pos=1/3] (sqp1_in1)
    coordinate[pos=2/3] (sqp1_in2)
};
\node [exbox, above=of t4, minimum width=2.25em] (sqp2) {SQP};
\path (sqp2.north west)--(sqp2.north east) {
    coordinate[pos=1/3] (sqp2_in1)
    coordinate[pos=2/3] (sqp2_in2)
};

\node [box, above=of t5, minimum height=2em, minimum width=3em] (pred) {$\cramped{\Model^{\ttt}_{\text{Pred}}}$};
\path (pred.south west)--(pred.south east) {
    coordinate[pos=1/5] (pred_out1)
    coordinate[pos=2/5] (pred_out2)
    coordinate[pos=3/5] (pred_out3)
    coordinate[pos=4/5] (pred_out4)
};
\draw[brace] ($(t5.south -| pred_out1)!-4pt!(t5.south -| pred_out4)$) -- ($(t5.south -| pred_out4)!-4pt!(t5.south -| pred_out1)$) node[bracenode, midway, below=4pt] (ylbl) {predicted $\hat{y}$};

\node [circle, draw, fill=lightblue, inner sep=0.5pt] at (t2.north |- sub.east) (sum) {$+$};
\node [exbox, above=1em of pred, minimum width=3.3em, text height=1.5ex, text depth=0.5ex] (filt) {$\sigma_{\scriptscriptstyle\!G} > \tau$};
\node [box, right=2.25em of sum, minimum height=1.75em] (se) {$\Model_{\text{SE}}$};
\path (se.north east)--(se.south east) {
    coordinate[pos=1/4] (se_out1)
    coordinate[pos=3/4] (se_out2)
};

\draw [ar] (x1.east) -- (sub.west);
\draw [ar] (sub.east) -- (mu.west |- sub.east) node[arnode] {noises};
\draw [ar] (s1.east)  -- (mu.west |- s1.east);
\draw [ar] (m1.east)  -- (mu.west |- m1.east);
\draw [arw] (s1.east) -| (sub.south);
\draw [ar]  (s1.east) -| (sub.south);
\draw [ar] (mu_out1.east) -- (sum.west) node[arnode] {noise $n$};

\draw [ar] (se_out1.east) -| (filt.north) node[arnode] {masks $G$};

\draw [ar] (mu_out3.east) -| (t1.north) node[arnode] {labels};
\draw [arw] (mu_out2.east) -| (sqp1_in1.north);
\draw [ar]  (mu_out2.east) -| (sqp1_in1.north);
\draw [ar] (mu_out2.east) -| (sqp2_in1.north);
\draw [arw] (mu_out2.east) -| (vad.north);
\draw [ar]  (mu_out2.east) -| (vad.north) node[arnode] {clean $s$};
\draw [arw] (mu_out2.east) -| (sum.south);
\draw [ar]  (mu_out2.east) -| (sum.south);
\draw [ar] (sum.east) -- (se.west) node[arnode] {noisy $x$};

\draw [ar] (se_out2.east) -| (sqp2_in2.north) node[arnode] {enhanced $\hat{s}$};
\draw [ar] (filt.south) -- (pred.north) node[arnode, below right] {masks $\tilde{G}$};

\draw [ar] (vad.south) -- (t2.north);
\draw [ar] (sqp1.south) -- (t3.north);
\draw [ar] (sqp2.south) -- (t4.north);
\draw [ar] (pred_out1.south) -- (t5.north -| pred_out1.south);
\draw [ar] (pred_out2.south) -- (t5.north -| pred_out2.south);
\draw [ar] (pred_out3.south) -- (t5.north -| pred_out3.south);
\draw [ar] (pred_out4.south) -- (t5.north -| pred_out4.south);

\draw [arw, rounded corners=1pt] (sum.east) -- ++(1em, 0em) coordinate (ext_start) -- (mu_out2.east -| sqp1_in2.north) coordinate (ext_mid) -- (sqp1_in2.north) coordinate (ext_end);
\draw [arw, rounded corners=1pt, ultra thick] ($(ext_start)!0.5!(ext_mid)$) -- (ext_mid) -- ($(ext_mid)!0.5!(ext_end)$);
\draw [ar, rounded corners=1pt]  (sum.east) -- (ext_start) -- (ext_mid) -- (ext_end);

\end{tikzpicture}}%
    \caption{Overview of the proposed system, showing data generation pipeline, \gls{se} model, targets extraction, and prediction models.}
    \label{fig:overview}
\end{figure}

\begin{figure*}[t]
    \centering
    \includegraphics[height=3.04cm]{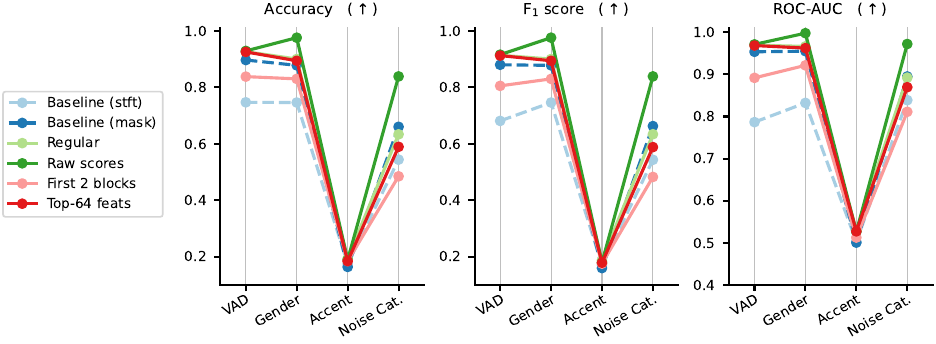}%
    \hfill%
    \includegraphics[height=3.04cm]{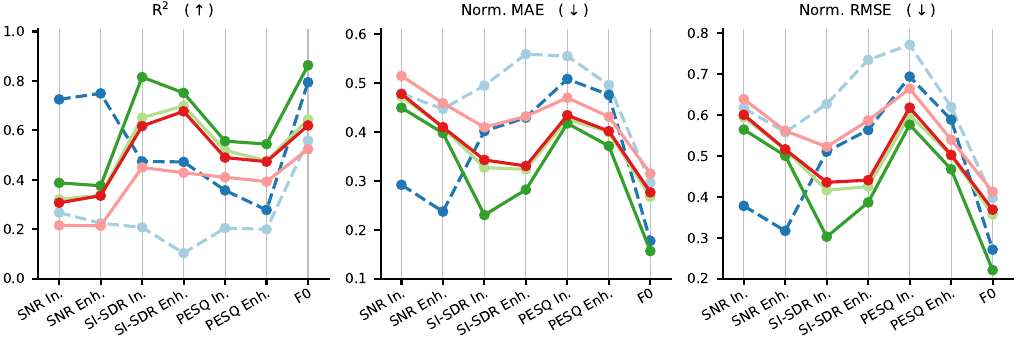}%
    \caption{Performance on each task using different input features (colors); first \num{3} subplots show classification, last \num{3} subplots show regression.}
    \label{fig:barplot}
\end{figure*}

\begin{figure}[t]
    \centering
    \includegraphics[width=1.0\linewidth]{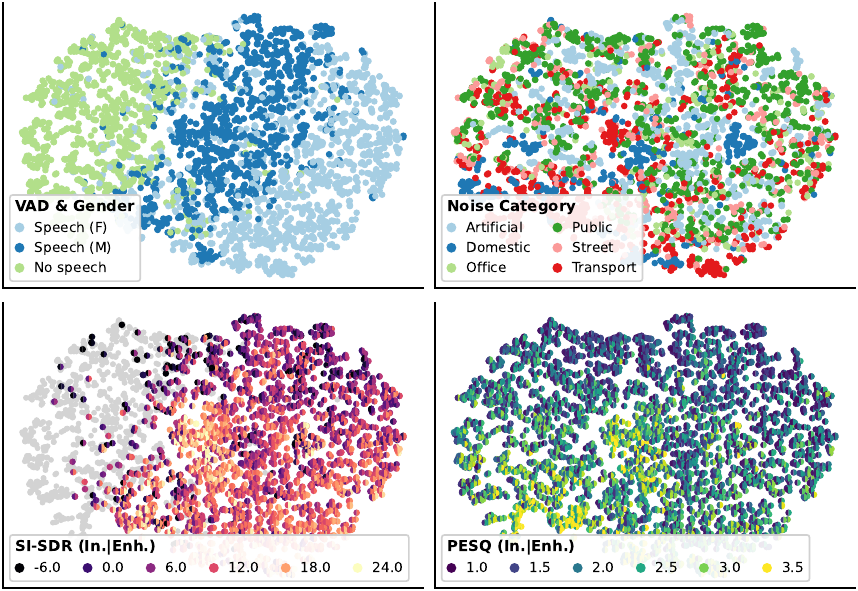}%
    \caption{Low-dimensional visualization of pruning masks, computed using t-SNE; for each subplot, points are colored by different targets.}
    \label{fig:tsne}
\end{figure}

The proposed system is illustrated in \cref{fig:overview}.
At its core, our work involves estimating $\smash{\cramped{y_{l}^{\ttt}} \; \forall \; t \in \mathbb{T}}$, where $\mathbb{T}$ is the set of prediction targets%
\footnote{Here, we use ``target'' for a given signal characteristic or class label and ``task'' for the associated prediction problem, often comprising multiple targets. %
Since linear model coefficients can be concatenated into a single matrix, we treat the two terms somewhat interchangeably.} %
(summarized in \cref{tab:targets}) and $l$ represents the current time step.
Since the \gls{se} model considered here operates in the \gls{stft} domain, each time step $l$ corresponds to a frame of window length $w$ and hop size $h$.
Accounting for the time dimension, our problem becomes:
\begin{equation}\label{eq:tasks}
    \cramped{y^{\ttt}} \approx \cramped{\hat{y}^{\ttt}} = \cramped{\Model^{\ttt}_{\text{Pred}}}(\tilde{G}); \quad \cramped{y^{\ttt}} \in \begin{cases}
        \cramped{\Real^L} & \text{\footnotesize{regression}} \\
        \cramped{\Bin^L} & \text{\footnotesize{classification}}
    \end{cases}
\end{equation}
where $L$ is length of the input sequence, $\cramped{\Model^{\ttt}_{\text{Pred}}}$ is the model associated with target $t$, consisting of a set of coefficients, and $\smash{\tilde{G}}$ is a matrix containing the pruning masks to be used as input features.

\subsection{Data Pipeline}
\label{ssec:data}

\gls{se} models are usually trained on batches of clean-noisy pairs, often corresponding to single utterances.
Since we are interested in real-time streaming use cases, we treat our training and test data as one long continuous signal.
As shown in \cref{fig:overview}, this is achieved by first extracting the individual noise excerpts from the noisy utterances.
Subsequently, we generate the clean speech and noise signals $s$ and $n$ by picking utterances/excerpts in random order.
Lastly, we mix them to form the noisy signal $x$.
The noisy mixture $x$ is then processed by the \gls{dyncp} \gls{se} model, yielding the enhanced speech signal $\hat{s}$.
To ensure a fair evaluation and avoid data leakage, we assign different speakers to the train and test splits and sample individual utterances using a stratified scheme that maintains a uniform distribution of gender and accent classes; we apply the same stratification strategy to the noise categories.
The audio signals are further processed to extract the ground truths for the regression tasks, while the binary pruning masks are gathered from the internal gating subnets.
See \cref{fig:data_example} for an example of the aforementioned generated data.

\subsection{Speech Enhancement with Dynamic Channel Pruning}
\label{ssec:dyncp}

We employ Conv-FSENet, introduced in \cite{miccini_scalable_2025}, as our \gls{dyncp} model for \gls{se}.
It is a \gls{stft}-domain model comprising $\smash{\cramped{\{\mathcal{B}_i\}_{i=1}^I}}$  processing blocks chained in series, each governed by a separate gating subnet.
The primary output of the model is the suppression mask $\smash{\hat{M}}$, used to recover the enhanced speech spectrum $\smash{\hat{S}}$ from noisy mixture $X$:
\begin{equation}
    S_{l,f} \approx \hat{S}_{l,f} = \hat{M}_{l,f} X_{l,f}; \quad \hat{M} \in \cramped{\Real^{L \times F}}\!; \quad X,S,\hat{S} \in \cramped{\Complex^{L \times F}}
\end{equation}
where $L$ and $F$, indexed by $l$ and $f$, are the number of time steps and frequency bins, respectively.
The model additionally produces a tensor of binary pruning masks $G$ determining which of the $\smash{\cramped{C_{\text{res}}}}$ convolutional channels in each block are active.
Thus, we have:
\begin{equation}
    \hat{M}, G = \Model_{\text{SE}}(X); \qquad G \in \cramped{\Bin^{L \times I \times C_{\text{res}}}}
\end{equation}

Although $G$ contains too much data to be viable as input features, not all of its masks are equally useful.
In fact, during training, the gating subnets might learn to treat certain channels as always needed or always unnecessary.
As a result, some of the masks in $G$ are constant or quasi-constant; such features would obviously be uninformative, so we filter them out by enforcing a minimum standard deviation $\tau$, resulting in a subset of $\smash{\cramped{C^{\smash{\star}}}} \leq \smash{\cramped{I C_{\text{res}}}}$ features:
\begin{equation}
    \tilde{G} \coloneqq \left\{ G_{:,i,c} \; \middle| \; \sigma_{G_{:,i,c}} > \tau \right\}; \qquad \tilde{G} \in \cramped{\Bin^{L \times C^{\star}}}
\end{equation}

\subsection{Targets Extraction}
\label{ssec:extract}

The ground truths for the prediction tasks, listed in \cref{tab:targets}, can be divided into discrete and continuous.
Some of the discrete targets are based on dataset metadata; when sampling utterances and noises to form $s$ and $x$, we also retrieve their corresponding metadata and append them to a time series whose time resolution matches that of the \gls{stft} frames. 
To estimate the \gls{vad} target, we: 
\begin{inlineenum_123}
    \item compute the \gls{rms} for each window;
    \item apply a zero-phase simple moving average filter;
    \item binarize the resulting energy envelope signal according to a threshold;
    \item smooth and binarize the coarse estimate again to obtain the final ground truth.
\end{inlineenum_123}

The continuous regression targets comprise speech quality metrics extracted from both noisy and enhanced signals using the \gls{sqp} modules shown in \cref{fig:overview}; these metrics are: 
\begin{inlineenum_abc}
    \item \gls{snr}, computed per-frame using the aforementioned \gls{rms} signals;
    \item \gls{sisdr}~\cite{le_roux_sdr_2019}, computed on \qty{1}{\second} windows with \qty{75}{\percent} overlap;
    \item \gls{pesq}~\cite{rix_perceptual_2001}, computed on \qty{3}{\second} windows with \qty{75}{\percent} overlap, padded with \qty{0.5}{\second} of silence on both ends;
\end{inlineenum_abc}
\gls{snr} and \gls{sisdr} values are converted to dB and clamped between \qty{-50}{\decibel} and \qty{30}{\decibel}.
To match the time resolution of the other ground truths, we upsample the \gls{sisdr} and \gls{pesq} targets to the \gls{stft} frame rate by interpolating them using a squared-Hann window.
We also extract the fundamental frequency $F_0$ of the clean speech using the WORLD vocoder~\cite{morise_world_2016}, with a frame period matching the \gls{stft} hop size $h$.

\section{Experimental Setup}
\label{sec:setup}

\begin{noindent_descr}
\item[Data generation]%
We use the speech utterances and noise excerpts from \gls{vbd}~\cite{valentini-botinhao_investigating_2016}, combining all of its \num{88} English and non-English speakers.
Since the speech data in \gls{vbd} is taken from \gls{vb}~\cite{veaux_voice_2013,yamagishi_cstr_2019}, we use its metadata for gender and accent labels.
Noise classes are based on the noise types reported in \gls{vbd} and matched with the categories listed in DEMAND~\cite{thiemann_demand:_2013}; an additional ``Artificial'' class is used to describe custom noises.
We keep the \num{5} most common accent classes and merge the rest into class ``Other''.
We retain the original gain of the noise excerpts, which offered a sufficient level of \gls{snr} variability.
We withheld \num{27} speakers for testing, i.e., \qty{\sim 30}{\percent} of the total.
This resulted in \qty{30}{\minute} of training data (\num{600} utterances, for a total of \num{112563} datapoints) and \qty{30}{\minute} of testing data (\num{594} utterances, totaling \num{112716} datapoints).
The \gls{se} backbone is parameterized with $C_{\text{res}} = 128$ convolutional channels and $I = 9$, i.e., \num{3} stacks with \num{3} blocks each.
The \gls{dyncp} gating subnets feature \num{16} hidden channels and pooling over the entire receptive field.
The model is trained as described in \cite{miccini_scalable_2025} with \num{0.25} target utilization and surrogate gradients.
We apply a threshold $\tau = 0.005$ to the standard deviation of the masks, resulting in $\smash{\cramped{C^{\smash{\star}}}} = 202$ features (\qty{\sim 18}{\percent} of all channels).

\item[Alternative features]%
We included the noisy input \gls{stft} log-magnitude and the predicted suppression mask $\smash{\hat{M}}$ as baselines; both comprise \num{257} features.
As additional experiments, we also trained the prediction models on: 
\begin{inlineenum_abc}
    \item features from the first \num{2} blocks only (\num{67} in total);
    \item top-\num{64} most informative features based on their linear coefficients on the regular setup;
    \item pruning scores $\smash{\tilde{R}}$, i.e., the raw outputs of the gating subnets before binarization; for consistency, we keep the same channels and dimensionality as $\smash{\tilde{G}}$.
\end{inlineenum_abc}
We apply z-score normalization to any non-binary variant, i.e., input spectrograms, predicted suppression masks, and raw pruning scores.

\item[Prediction models]%
We employ logistic and linear regression models for classification and regression tasks, respectively.
We apply $\smash{\ell_2}$ (Tikhonov) regularization with a factor $\alpha = 0.01$ to account for feature collinearity.
We evaluate classification performance by accuracy, $\smash{\cramped{F_1}}$-score, and \gls{rocauc}.
For regression, we use \gls{r2}, \gls{mae}, and \gls{rmse}; to simplify comparison across tasks, we normalize the two latter metrics by the interquartile ranges shown in \cref{tab:targets}.
Both metrics and models rely on the \texttt{scikit-learn} library.
Lastly, gender/accent classification and \gls{snr}/\gls{sisdr} regression are trained and evaluated only on frames with voice activity (\qty{\sim 68}{\percent} of all data), using the \gls{vad} target as oracle.
Similarly, fundamental frequency estimation only employs frames where $F_0 \neq 0$.

\item[\acrfull{sv}]%
We explore the viability of pruning masks as embeddings for \gls{sv}.
For each utterance $u \in \mathcal{U}$, we define $\smash{\cramped{\tilde{G}^{\smash{(u \land v)}}}}$ as the subset of mask frames from utterance $u$ featuring voice activity, implied by $v$.
Each subsequence is then averaged and $\smash{\cramped{\ell_2}}$-normalized, yielding the set $\mathcal{E}$ of utterance-level embeddings $e$:
\begin{equation}
    \mathcal{E} \coloneqq \left\{ \frac{\cramped{\bar{G}^{\uuu}}}{\norm{\cramped{\bar{G}^{\uuu}}}_2} \; \middle| \; u \in \mathcal{U} \right\}; \qquad \cramped{\bar{G}^{\uuu}} = \mathbb{E}_{l} \left[\cramped{\tilde{G}^{(u \land v)}}\right]
\end{equation}
where $\mathcal{U}$ is the set of all utterances and $\smash{\cramped{\bar{G}^{\uuu}}}$ is the average across time of each aforementioned subsequence.
We employ a standard linear backend consisting of \gls{wccn} and \gls{lda} (with \num{16} output dimensions) followed by length normalization and cosine similarity for scoring, similar to \cite{dehak_front-end_2011}.
We fit the backend on the training split and evaluate on the test data.
For each speaker, we assign $\smash{\cramped{N_{\text{enr}} \in \{1, 2, 3\}}}$ embeddings from $\mathcal{E}$ as enrollment (averaging them when $\smash{\cramped{N_{\text{enr}} > 1}}$) and perform verification trials on the remaining ones. 
We generate pairs of enrollment and test utterances with a \numrange[range-phrase=:]{1}{10} ratio of target and non-target speakers, totaling between \num{5670} and \num{5130} trials depending on $\smash{\cramped{N_{\text{enr}}}}$.
Lastly, we omit the suppression mask baseline and instead show the enhanced \gls{stft} log-magnitude and the top\nobreakdashes-\num{64} most informative raw pruning scores.
As customary for \gls{sv}, we evaluate performance using the \gls{eer}.

\end{noindent_descr}

\section{Results}
\label{sec:results}

\begin{figure}[t]
    \centering
    \includegraphics[width=1.0\linewidth]{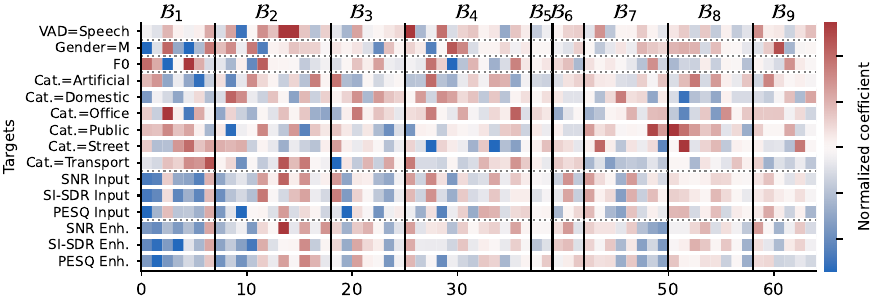}%
    \caption{Normalized coefficients (red for positive, blue for negative) for models trained on the top\nobreakdashes-\num{64} most informative features (x-axis, grouped by processing block); showing a subset of targets (y-axis).}
    \label{fig:heatmap}
\end{figure}




\begin{figure}[t]
    \centering
    \includegraphics[width=1.0\linewidth]{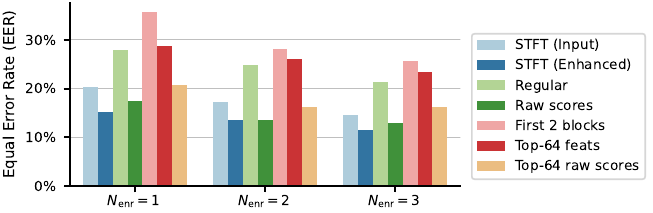}%
    \caption{\gls{sv} performance across different number of enrollment utterances (x-axis) and feature sets (colors).}
    \label{fig:sv_results}
\end{figure}

\cref{fig:barplot} compares predictors trained on features from \gls{dyncp} against baseline data. 
Regular binary masks (\colordot{lightgreen}) outperform both baselines across most classification and regression tasks, with the 64 most informative features (\colordot{red}) retaining the same performance as the full set (\colordot{lightgreen}), suggesting that auxiliary models can be very compact.
Conversely, only using data from the first two blocks (\colordot{lightred}) causes a significant performance drop, potentially implying that the gating subnets learn progressively more complex features throughout the blocks.
Raw scores (\colordot{green}) are strongest on nearly all tasks, with the highest gains in noise classification and input \gls{sisdr}/$F_0$ estimation; this is expected, as they carry more information about the internal state of the \gls{se} model.
Accent is remarkably hard to predict, with near-chance performance across all variants, plausibly because \gls{se} behavior is not particularly affected by this factor. 
Lastly, the $\smash{\hat{M}}$ baseline (\colordot{blue}) dominates tasks with rapidly varying targets such as \gls{snr} or $F_0$.
In these cases, features from \gls{dyncp} models are inherently penalized, due to the temporal averaging occurring inside the gating subnets. 

Using the top\nobreakdashes-\num{32} principal components of \qty{20}{\percent} of the binary pruning masks in the test set, we fit a t-SNE manifold to visualize them in a low-dimensional space.
As evidenced in \cref{fig:tsne}, the signal characteristics are arranged in a semantically meaningful way.
In particular, we observe distinct clusters for voice activity, with secondary separation by speaker gender and continuous gradients for \gls{sisdr} and \gls{pesq}. 
Interestingly, a closer inspection of input and enhanced labels reveals slightly different trajectories between targets, with these trends extending into noise-only regions in the case of \gls{pesq}.
Somewhat expectedly, noise categories fragment into smaller, scattered clusters: indeed, the metadata used as labels mostly describe the recording environment rather than the actual noise content.

The heatmap in \cref{fig:heatmap} shows how each pruning mask contributes to the estimation of the different targets.
It comprises coefficients of models trained on the top\nobreakdashes-\num{64} binary features (\colordot{red} in \cref{fig:barplot}).
To simplify comparison across different tasks and value ranges, we normalize them by first multiplying them by the standard deviation of their respective input features and then scaling to unit $\smash{\ell_2}$-norm.
Once again, several trends emerge; intuitively, features driving male classification overlap with those used for fundamental frequency regression, albeit with opposite polarity, consistent with known gender-pitch correlation.
Conversely, in noise category classification, classes rely on disjoint subsets of features, with the few shared ones having different polarity.
Meanwhile, \gls{snr}, \gls{sisdr}, and \gls{pesq} --- although to a lesser extent --- rely on overlapping sets of features, with many belonging to the earlier blocks and featuring negative coefficients.
These observations may be taken as evidence that the underlying \gls{dyncp} \gls{se} model learns to behave conservatively, only activating the channels associated with those masks when the input signal is particularly noisy.
Lastly, and rather counterintuitively, some \gls{vad}/gender-relevant features also appear to inform noise classification, which could indicate occasional confusion between background and target speech.

The results of our last set of experiments, tackling \gls{sv}, are shown in \cref{fig:sv_results}.
Using our relatively simple setup, the \gls{dyncp} masks achieve only modest \gls{eer}.
This is particularly true for binary features (\colordot{lightgreen}, \colordot{lightred}, and \colordot{red}), mostly ranking worse than the \gls{stft} baselines (\colordot{lightblue} and \colordot{blue}).
Nevertheless, the full set of raw scores (\colordot{green}) performs similarly, while requiring \qty{21}{\percent} fewer operations; dropping the number of features to \num{64} (\colordot{lightorange}), however, heavily degrades performance.
It is unclear whether the lackluster results are due to information loss from temporal pooling, or because the underlying \gls{se} model learns an internal representation that is mostly speaker-agnostic, thereby abstracting away individual speaker characteristics.

\vfill

\section{Conclusion}
\label{sec:conclusion}



We showed that the binary pruning masks learned by a \gls{dyncp} \gls{se} model expose linearly accessible information about speech and acoustic properties of its input, hinting at local competition inside the model~\cite{srivastava_understanding_2015}.
With as few as \num{64} features, we achieve \qty{93}{\percent} accuracy on \gls{vad} and \qty{59}{\percent} on noise classification; when predicting input \gls{pesq} and \gls{sisdr}, we obtain a \gls{mae} of \num{0.2} and \qty{3.2}{\decibel}, respectively.
Most crucially, these auxiliary estimates come with almost negligible computational overhead; when considering all classification and regression tasks (totaling \num{21} targets) on the full set of \num{202} features, we only incur \num{4242} additional arithmetic operations per frame, i.e., somewhere between \qty{0.6}{\percent} and \qty{0.93}{\percent} of the total compute, depending on the number of currently active channels.

While far from mature, our preliminary results pave the way for \gls{dyncp}-based single-model pipelines capable of simultaneously enhancing speech, saving compute, and producing auxiliary data that can aid robustness and user experience.
Nonetheless, we point out the following limitations: 
\begin{inlineenum_abc}
    \item linear models assume additive contributions and are sensitive to correlated inputs;
    \item collinearity between features makes interpretation of model coefficients as feature \emph{importance} difficult and potentially misleading;
    \item temporal smoothing may hinder estimation of quickly-changing targets;
    \item the windowed \gls{pesq} ground truths adopted here might not be fully representative of the instantaneous signal quality.
\end{inlineenum_abc}
As part of our research, we plan to investigate other types of dynamic models, and experiment with simultaneously training on \gls{se} and auxiliary tasks.

\section{References}
\printbibliography[heading=none]

\end{document}